\documentclass[pdflatex,snbasic]{nature}

\usepackage{graphicx}%
\usepackage{multirow}%
\usepackage{amsmath,amssymb,amsfonts}%
\usepackage{amsthm}%
\usepackage{mathrsfs}%
\usepackage[title]{appendix}%
\usepackage{xcolor}%
\usepackage{textcomp}%
\usepackage{manyfoot}%
\usepackage{booktabs}%
\usepackage{algorithm}%
\usepackage{algorithmicx}%
\usepackage{algpseudocode}%
\usepackage{listings}%

\setcitestyle{numbers}
\setcitestyle{square}

\begin{document}

\title[Article Title]{Actionable Insights on Philadelphia Crime Hot-Spots: Clustering and Statistical Analysis to Inform Future Crime Legislation}

\author[1,2]{\fnm{Ishan S.} \sur{Khare}}
\equalcont{All authors contributed equally to this work.}

\author[1,2]{\fnm{Tarun} \sur{Martheswaran}}
\equalcont{All authors contributed equally to this work.}

\author[1,2]{\fnm{Rahul} \sur{Thomas}}
\equalcont{All authors contributed equally to this work.}

\author[1,3]{\fnm{Aditya} \sur{Bora}}
\equalcont{All authors contributed equally to this work.}

\affil*[1]{\orgdiv{Computer Science Department}, \orgname{Stanford University}, \orgaddress{\city{Stanford}, \country{USA}}}

\affil[2]{\orgdiv{Mathematics Department}, \orgname{Stanford University}, \orgaddress{\city{Stanford}, \country{USA}}}

\affil[3]{\orgdiv{Electrical Engineering Department}, \orgname{Stanford University}, \orgaddress{\city{Stanford}, \country{USA}}}

\maketitle

\newpage

\section{Abstract}
\label{sec:abstract}
Philadelphia's problem with high crime rates continues to be exacerbated as Philadelphia's residents, community leaders, and law enforcement officials struggle to address the root causes of the problem and make the city safer for all. In this work, we deeply understand crime in Philadelphia and offer novel insights for crime mitigation within the city. Open source crime data from 2012-2022 was obtained from OpenDataPhilly. Density-Based Spatial Clustering of Applications with Noise (DBSCAN) was used to cluster geographic locations of crimes. Clustering of crimes within each of 21 police districts was performed, and temporal changes in cluster distributions were analyzed to develop a Non-Systemic Index (NSI). Home Owners' Loan Corporation (HOLC) grades were tested for associations with clusters in police districts labeled `systemic.' Crimes within each district were highly clusterable, according to Hopkins' Mean Statistics. NSI proved to be a good measure of differentiating systemic ($<$ 0.06) and non-systemic ($\geq$ 0.06) districts. Two systemic districts, 19 and 25, were found to be significantly correlated with HOLC grade (p $=2.02 \times 10^{-19}$, p $=1.52 \times 10^{-13}$). Philadelphia crime data shows a high level of heterogeneity between districts. Classification of districts with NSI allows for targeted crime mitigation strategies. Policymakers can interpret this work as a guide to interventions. \vspace{0.3cm}

\textbf{Keywords:}
Patterns, Redlining, Clustering, Philadelphia
\maketitle
\section{Background and Introduction}\label{sec:intro}
Philadelphia is one of the most populated cities in the United States, with a rich history and a diverse population. However, like many large cities, Philadelphia has struggled with high crime rates for decades. It has long been known for its high crime rates, particularly violent crimes such as homicide and aggravated assault. In recent years the city continues to experience higher crime rates compared to many other large cities in the United States \cite{abrams2021covid}. According to the Philadelphia Police Department, in 2021 there were 499 homicides, which is the highest number in the city in 30 years \cite{philly_police_2022}.

Over the past 10 years, Philadelphia has implemented various policing strategies aimed at reducing crime rates in hotspot districts. These include community policing, which emphasizes collaboration and communication between the police and community members, as well as hot-spot policing, which focuses on targeting areas with high levels of criminal activity \cite{balocchi_2019}. Despite these efforts, crime rates in some districts of Philadelphia remain stubbornly high: policing these districts requires a nuanced understanding of the local community, its needs, and its priorities \cite{ratcliffe_2021}.

Indeed, one of the key challenges in addressing crime in Philadelphia is the issue of policing districts. Philadelphia is divided into 21 police districts, each of which has its own unique characteristics and challenges. Some districts, such as Center City and University City, are densely populated and experience a high volume of tourism and commerce, while others are more residential and may be marked by poverty and disinvestment \cite{ehlenz_2016}. Any crime prevention strategy that neglects these differences will ultimately be ineffective. 

Furthermore, it is critical that community leaders understand the policing alone is very unlikely to reduce crime \cite{dungy_2021}. Especially given the recent events regarding police brutality, is it paramount that policy makers and community leaders use different methods to try and reduce crime. For instance, tools such as educational investments and infrastructure investments have proven to be effective ways of solving root causes of crime \cite{henson_2022}. These among other techniques should be heavily considered and appropriately implemented when creating crime mitigation strategies \cite{montolio_2018_the}. 

Our project aims to explore and utilize the temporal and geospatial trends in Philadelphia's crime rates in order to better inform the policing, infrastructure, and educational decisions within particular districts. Ultimately, we seek to provide insight into the complex and multifaceted nature of crime and policing in Philadelphia. By examining the data collected regarding previous crimes, demographics, systemic discriminatory practices like redlining, among other factors, we formed a deeper understanding of the issues exacerbating crime rates in Philadelphia.

Ultimately, we discovered that crime can be classified based on location and temporal features which can assist policy makers in creating more effective targeted legislation.

\section{Technical Exposition}
\subsection{Data Cleaning and Feature Engineering}

We extensively use crime data from OpenDataPhilly, providing a comprehensive table of data of crimes reported in Philadelphia from 2012 to 2022 \cite{opendataphilly}. The particularly important features are the time of dispatch, geospatial data represented in terms of latitude and longitude, the street address of the crime, and associated police district. \textit{The main feature we use is the geospatial data, in terms of latitude and longitude.}

Data cleaning of crime data is kept relatively simple, as the main data we wish to use is the geospatial information of crimes. First, we delete rows with `NA' listed under the location. Second, we remove rows with invalid dispatch times and dates, and street addresses. These invalid entries account for $0.7\%$ of all rows. In addition, we remove data points belonging to districts 4 and 23 as there is only one datapoint for each of the districts.

Finally, using the district feature, we computed aggregate counts of crime in each district from 2012 to 2022. We found that in districts $4$ and $23$, there had only been one crime in $10$ years, so we removed these entries. This completes the cleaning and feature engineering for data obtained and compiled from the OpenDataPhilly dataset.

\subsection{Exploratory Data Analysis} 

\subsubsection{Understanding Police Districts}

In Philadelphia, a police district is a geographic area that is patrolled by officers from the Philadelphia Police Department. According to the City of Philadelphia, there are a total of 21 police districts in the city.

Each police district covers a specific area of the city and is responsible for responding to emergency calls, investigating crimes, and maintaining public safety in that area. The boundaries of the police districts are based on factors such as population density, crime rates, and geographic features. Each district is staffed by police officers, detectives, and support personnel and is responsible for responding to calls for service, conducting investigations, and engaging with the community \cite{city_website}.

Given that police resources are allocated within these districts, we begin our analysis by analyzing trends with respect to these district boundaries. We also choose to analyze crime distributions within Philadelphia without the context of the police district boundaries to see if there are larger conclusions/trends which cross district borders. 

\subsubsection{Temporal Analysis By Police District}

\begin{figure}[ht]
\begin{center}

\includegraphics[width=1.0\textwidth]{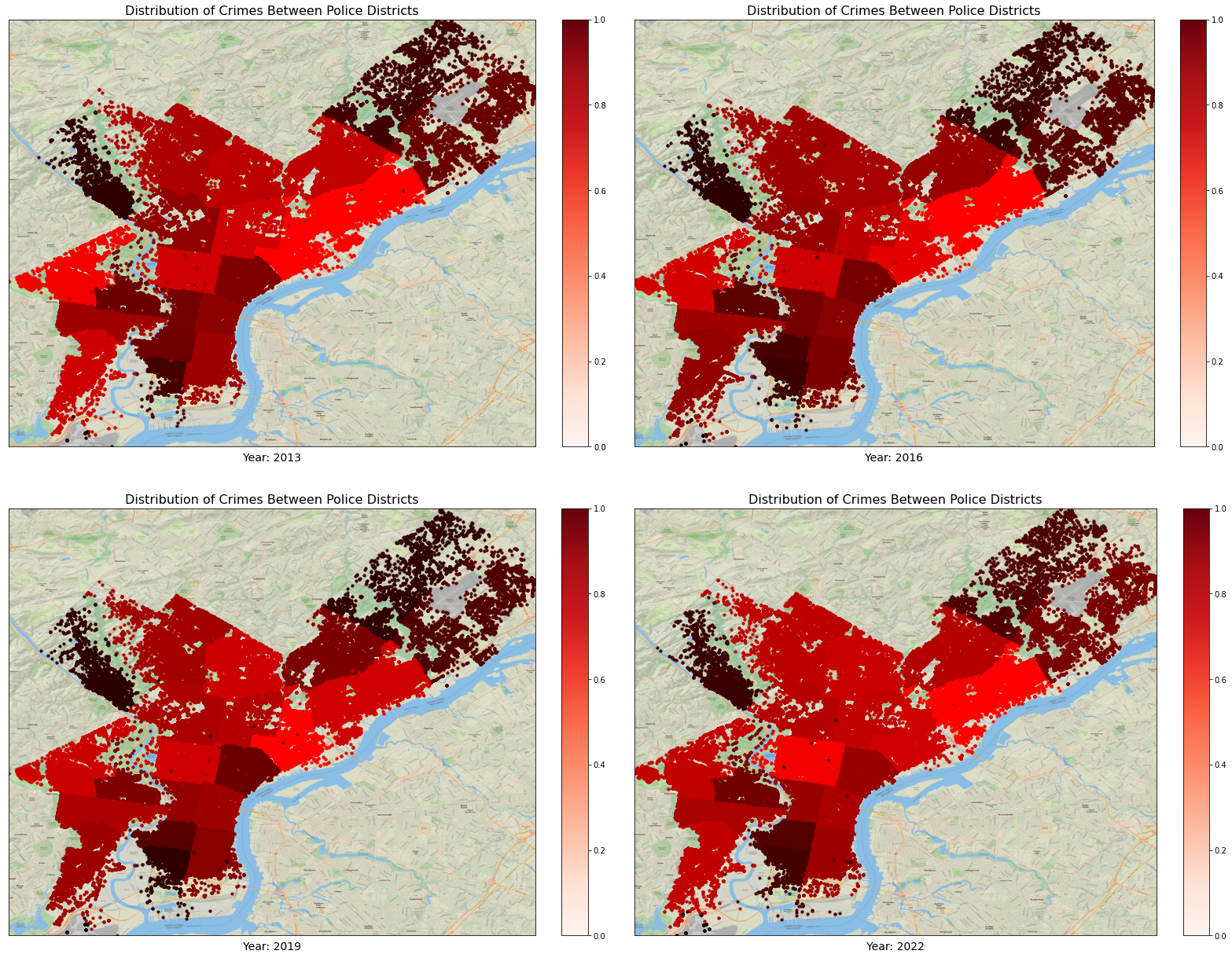}
\caption{Heat map of crime percentages for each Philadelphia police district from 2013 - 2019. Dark red regions indicate high crime frequency whereas light red regions indicate lower crime frequency.}
\label{fig:dist-heat}
\end{center}
\end{figure}

To better understand the city of Philadelphia's crime incidence we wish to granularize our analysis. An obvious point of exploration is a location based exploration, specifically by police district. With the upcoming plots we wish to reveal characteristics about the police district level crime rate. Specifically we want to determine the relative frequency of the crime per district. Here we create heatmaps of the crime distribution by police district across the city of Philadelphia over a 9-year time frame with 4-year time increments (Figure \ref{fig:dist-heat}). 

\begin{figure}[H]
\begin{center}
\includegraphics[width=1.0\textwidth]{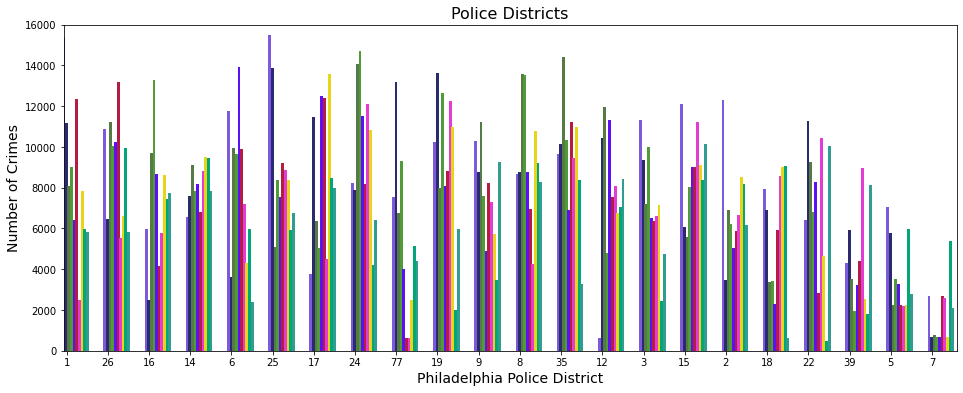}
\caption{For each Philadelphia police district, we plot the crime frequencies for every year from 2013 - 2019.}
\label{fig:temp_dist}
\end{center}
\end{figure}

At first glance, we see that there are no changes of greater than 0.5 with respect to the distribution over time within certain police districts. However, we do observe that there are general shifts in distribution within the districts which motivates a more narrowly-focused exploration into individual geographic regions (i.e. districts). This narrow-focus would allow for the discovery of any potential temporal changes that are not otherwise recognizable. In particular, Figure \ref{fig:dist-heat} and Figure \ref{fig:temp_dist} help us identify and segment out districts where the crime frequency remained consistent over time versus districts where the crime frequency changed. Examining districts individually can help us determine whether there was uniformity in crime frequency within a given district or not. Such exploration may be useful in determining if the general trend for each district casts a shadow upon or hides meaningful insight at a finer level.

\subsubsection{DBSCAN Clustering}

From the previous heat map visualizations in Figure \ref{fig:dist-heat}, we infer that we should \textit{cluster} crimes to extrapolate geospatial hotspots with similar levels of crime. One common clustering algorithm is \textit{$k$-means clustering}, where we group all points into $k$ clusters of close points. This is done by minimizing an objective function, based on sums of distances squared of points to the centroids of their clusters. However, there are a few downsides to this algorithm:
\begin{itemize}
    \item First, running $k$-means requires prior knowledge of the number of clusters, which is inconvenient for a large of clusters.
    \item Furthermore $k$-means has difficulty clustering data which has pockets of different sizes and densities, which is something we observe in our heat map visualizations.
    \item Finally, $k$-means performs poorly when a large number of outliers are present.
\end{itemize}

Given these obstacles, we instead opt for a clustering algorithm called \textbf{DBSCAN (density based spatial clustering)} \cite{ester1996density}. This is a non-parametric algorithm, which groups together points which are packed together; it does not required knowledge of the cluster count beforehand. It is also one of the most commonly-used clustering algorithms, and has been proven to be particularly well-suting in dealing with latitudes and longitudes. Thus, it is well-suited to the task of clustering crimes based on geospatial proximity. Another advantage of DBSCAN is its ability to separate high density areas from low density areas. As seen in the heat map, crime percentages range from $0.4\%$ to over $1\%$, so we can infer DBSCAN will perform well in this regard.

The main idea behind DBSCAN is that a point belongs to a cluster if it is close to \textit{many} points from that cluster. There are two parameters used in DBSCAN: $$\epsilon: \text{minimum distance away for a new point to be added to a cluster,}$$ and $$p_{min}: \text{minimum number of data points to define a cluster.}$$
To choose $\epsilon$, we use KNN ($k$ nearest neighbors) \cite{knn_1952}. By plotting the geospatial distances from one data point to the others, sorted in ascending order, we form an ``elbow'' curve. This curve has a sharp change in concavity, which is where we choose $\epsilon$. The value of $p_{min}$ is traditionally chosen to be twice the number of features, so in this case, we set it to $2 \cdot 2 = 4$. 

Once DBSCAN is finished running, it classifies all points into three levels of ``density''. 
\begin{itemize}
    \item A \textit{core point} is one where at least $p_{min}$ points within distance $\epsilon$.
    \item A \textit{border point} is one that lies at most $\epsilon$ away from a core point, but less than $p_{min}$ points within distance $\epsilon$.
    \item An \textit{outlier (noise)} is anything that is not a core or border point.
\end{itemize}
The clusters we obtain will consist of the core points: these represent crime hotspots, areas where the geospatial distribution of crimes for a particular year is particularly dense.

\subsubsection{Clustering Within Philadelphia Over Time}

\begin{figure}[H]
\begin{center}
\includegraphics[scale=0.4]{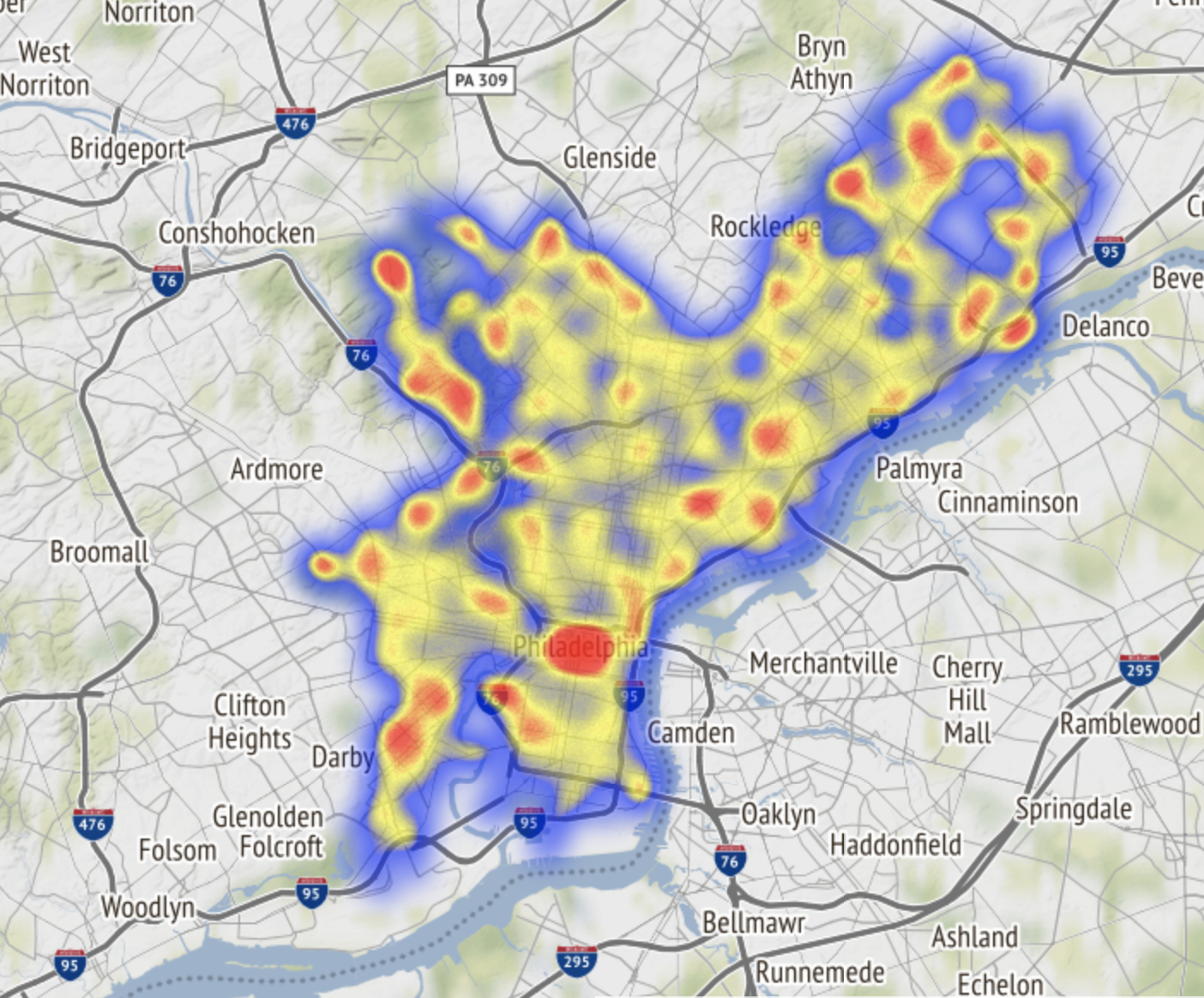}
\caption{This figure shows a heatmap of crime frequency over all of Philadelphia for the years 2012-2022.}
\label{fig:heatmap-all}
\end{center}
\end{figure}

Our preliminary motivation for this section is to analyze crime within Philadelphia without the context of the police borders. The reason for this is to better understand trends in crime that cross the boundaries. Initially, we plotted a heatmap of crime frequency over all of Philadelphia for the years 2012-2022, as displayed in Figure \ref{fig:heatmap-all}. We can visually observe that there seem to be pockets of higher crime frequency, which motives the need for clustering. Clustering can help identify spatial patterns and groupings in geographic data, allowing for more efficient analysis and decision-making.

Here, we have used DBSCAN clustering over a $9$-year time frame. Above, we have selected clustering graphs for every $3$ years, starting at $2013$ and ending at $2022$. Observe that the the locations of particular clusters change significantly over time. Also, notably, the number of clusters has decreased over time. Furthermore, when we overlay one such cluster plot with the district lines, we observe that clusters overlap between different districts. This is actually a reason to consider \textit{clustering within each district, rather than on all Philadelphia:} this overlap suggests an extra level of complexity that cannot be resolved by solely district or solely proximity of latitude and longitude, so we are motivated to use a combination of both.

\begin{figure}[H]
\begin{center}
\includegraphics[width=1.0\textwidth]{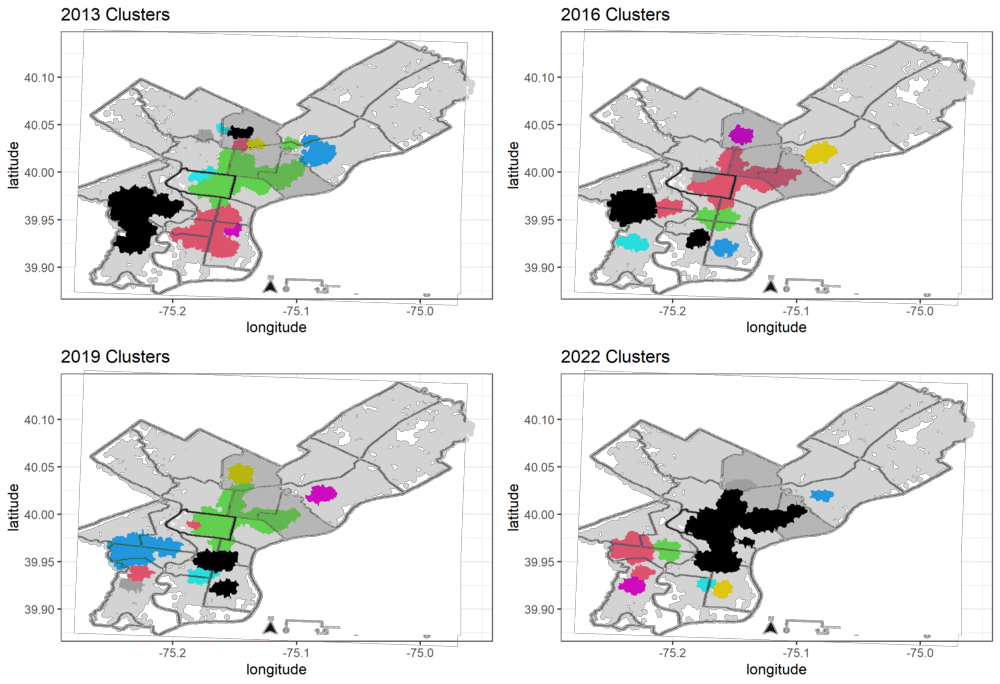}
\caption{DBSCAN's Clustering over 9-year time frame}
\label{fig:dbscan-all}
\end{center}
\end{figure} 

Now, there are a few other reasons that we want to run DBSCAN clustering within each district. First, from our qualitative observations in Figure \ref{fig:dbscan-all}, the density is different within each district, so it would make sense to represent this non-uniformity through district-focused clustering. Second, aggregate clustering gives at most $13$ clusters in all cases, which is significantly less than the number of actual police districts: so, we might gain better insight into geospatial trends of crime by specifying the algorithm runs within each district. Finally, with regards to politics, we would expect districts to separate their police officers and resources: so any sort of policing strategy \textit{should not operate based on clusters which lie in multiple districts.}

\subsubsection{Hopkins statistic testing}

Before we run clustering on each district, we introduce a statistical testing method which evaluates the success of clustering. The Hopkins statistic provides a measure of how "uniform" a collection $\mathcal{P}$ of $N$ points is \cite{hopkins}. When the Hopkins value is close to $0$, the data is largely uniform, and when it is close to $1$, the data is highly clustered. 

We compute the Hopkins statistic as follows. We first select $n << N$, which will be our subcollection size for the testing. Then, we choose a random subset $\mathcal{Q}$ of $n$ points from $\mathcal{P}$. Next, we randomly and uniformly generate a set $\overline{\mathcal{Q}}$ of $n$ random points. Finally, as our data is $2$-dimensional, the Hopkins statistics for these random sets $\mathcal{Q},\overline{\mathcal{Q}}$ are given by
\begin{equation}H(\mathcal{Q},\overline{\mathcal{Q}}) = \frac{\sum_{q \in \mathcal{Q}} \min_{p \neq q \in \mathcal{P}} d(p,q)^2}{\sum_{q \in \mathcal{Q}} \min_{p \neq q \in \mathcal{P}} d(p,q)^2 + \sum_{\overline{q} \in \overline{\mathcal{Q}}} \min_{p \in \mathcal{P}} d(p,\overline{q})^2}.\end{equation}
Here, $d(p,q)$ is the distance between $p$ and $q$.

\subsubsection{Clustering within districts over time}

\begin{figure}[ht]
\begin{center}
\includegraphics[scale=0.55]{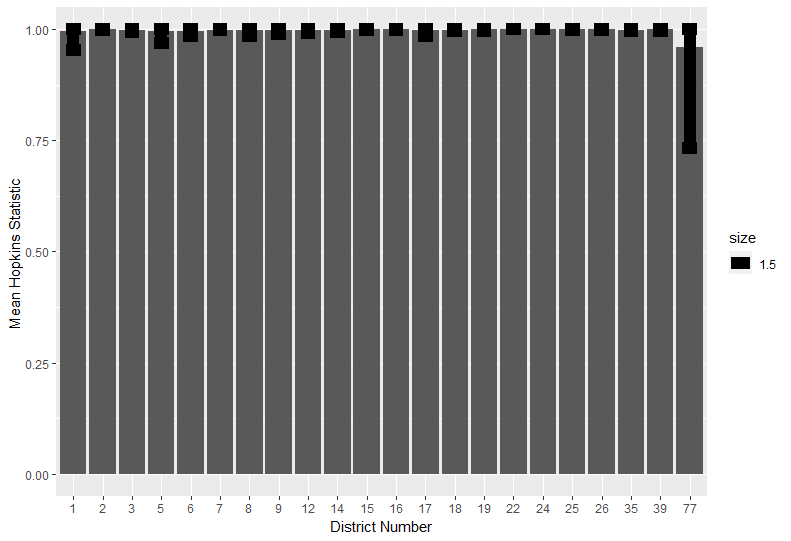}
\caption{Mean Hopkins Statistic Per District}
\label{fig:hopkins}
\end{center}
\end{figure} 

The results of the Hopkins test, as displayed in Figure \ref{fig:hopkins} show that  within each district the data is highly cluster-able. This justifies the use of DBScan to visualize these clusters and determine their geographic displacements over time which is critical as explained in methodology.  

\subsection{Methodology}

The Hopkins test showed us that the crime distribution complexity could be well-modeled within each of the police districts. Given this result, we propose the following methodology to help classify the type of crime occurring in a district in order to implement prevention and mitigation measures which are most likely to be successful.

The overarching goal is to be able to create a model which helps policy makers develop the \textbf{most targeted interventions to help prevent crime}.

\subsection{Defining "systemic" vs "non-systemic" crime}

\begin{figure}[ht]
\begin{center}
\includegraphics[scale=0.45]{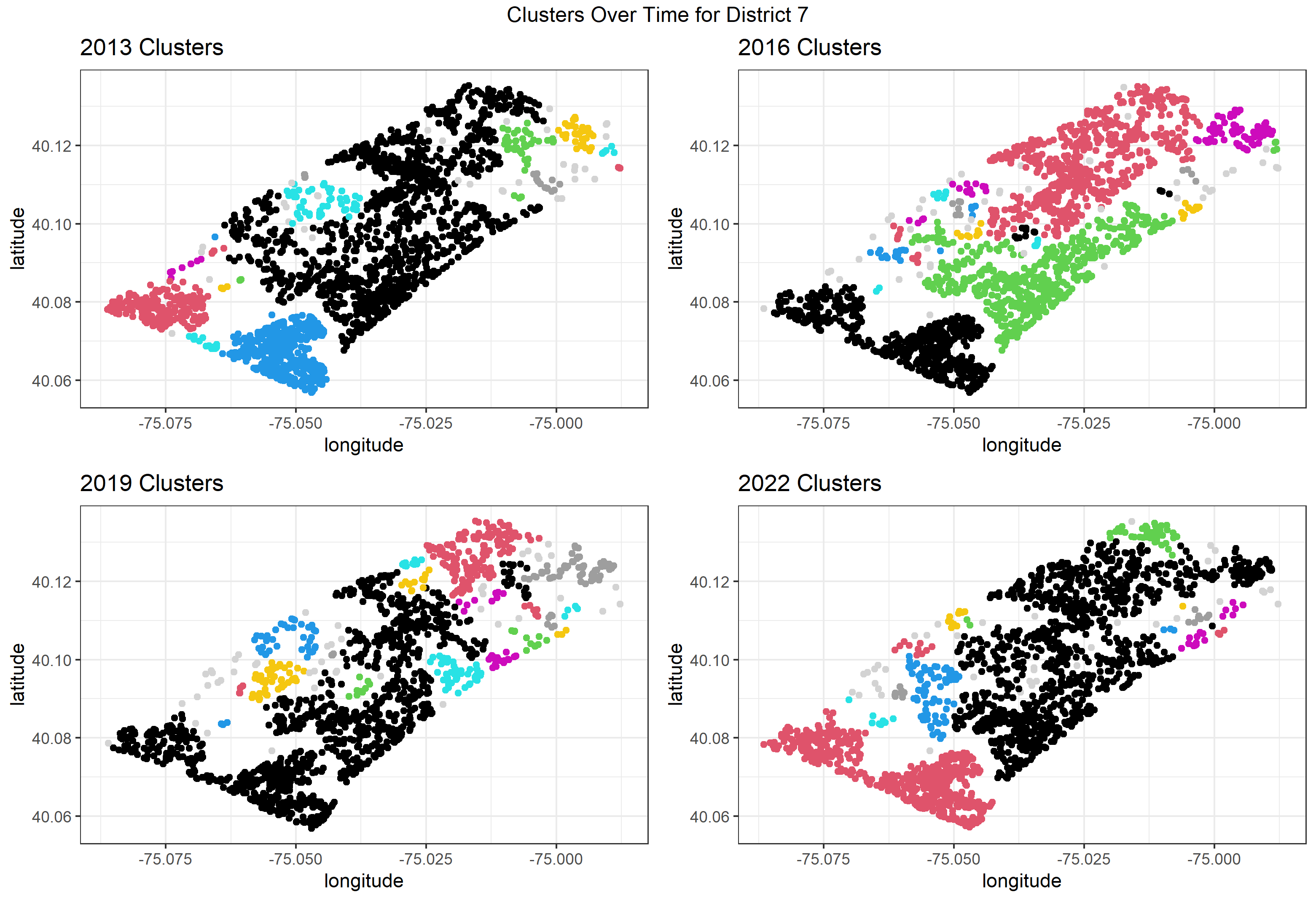}
\caption{District 7: Clusters can be seen to change geographic location over time}
\label{fig:District7}
\end{center}
\end{figure}

\begin{figure}[ht]
\begin{center}
\includegraphics[scale=0.45]{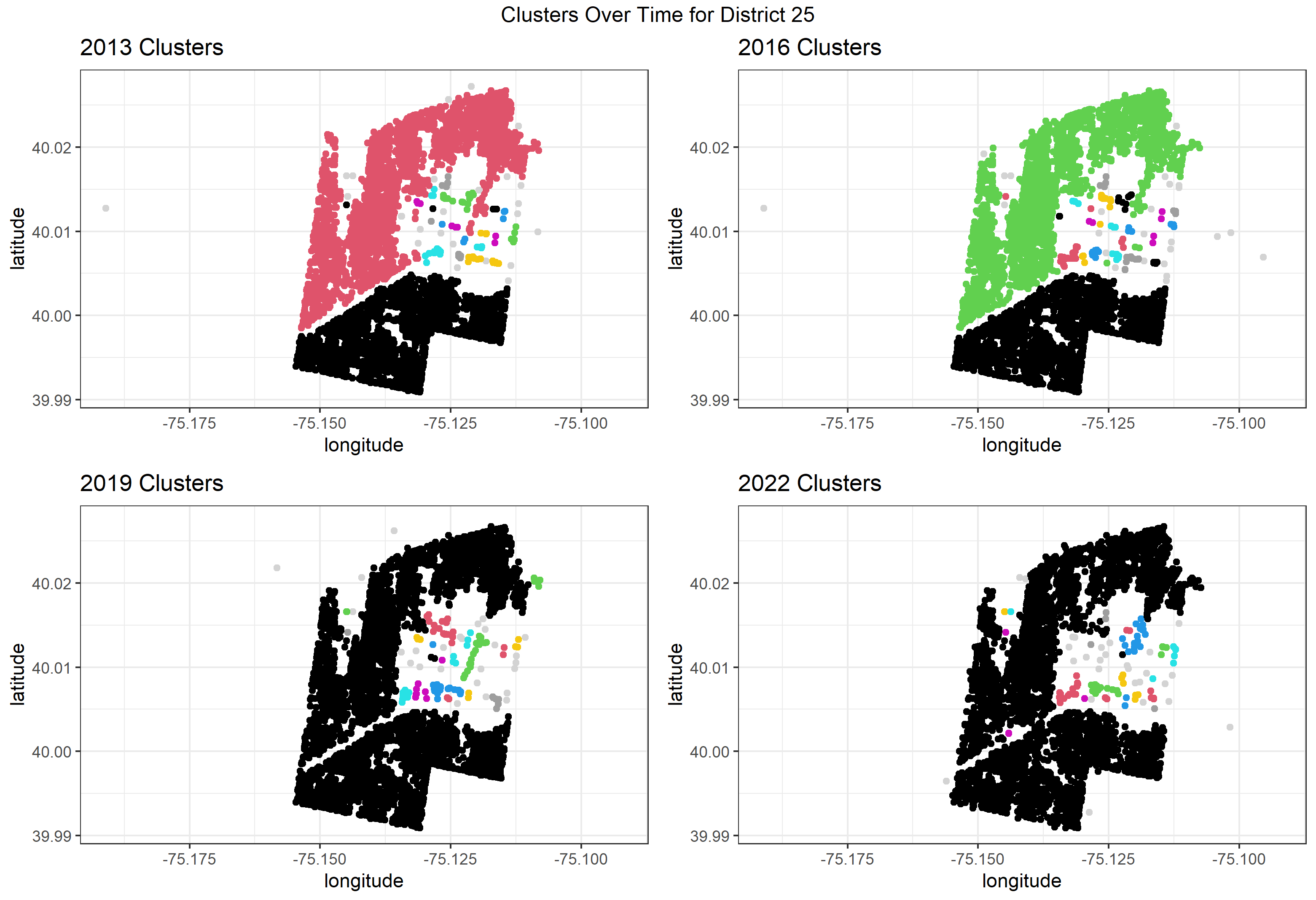}
\caption{District 25: Clusters can be seen to stay in the same geographic location over time}
\label{fig:District25}
\end{center}
\end{figure}

After visual analysis of the DBSCAN output on the 21 police districts, a clear theme could be discerned: certain districts would have clusters which don't change geographically over time (Figure \ref{fig:District25}) whereas other districts do (Figure \ref{fig:District7}). We believe that differentiating these districts can be extremely useful for policy makers to implement targeted solutions to try and reduce crime. 

Specifically, we propose a method for classifying districts within two categories, systemic and non-systemic, which we define below. This classification allows for targeted implementation of interventions measures. 

\subsubsection{High Level Overview Of Systemic Crime}
We have defined systemic crime to be a district in which the crime clusters have not changed in geography over time (statistical procedure for this determination is below). Given this, we suggest that interventions which address the root causes of crime will be more effective than increased policing. For instance, a tool like education and infrastructure investments would be more fruitful interventions in these areas. We attempt to discern factors which drive the "systemic-ness" of the crime such as comparing against historic spatial racism which refers to the use of geographic or spatial factors to perpetuate discriminatory practices.  

\subsubsection{High Level Overview Of Non-Systemic Crime}
We define a Non-systemic crime district to be one in which the clusters of crime changed geographic locations over time. A real world example of a non-systemic crime district could be one in which crime spiked as a result of recent civil unrest. In a situation such as this, data  suggests that increased policing would be more effective in decreasing the rate of crime given that the crime is "non-systemic" in nature \cite{AtlantaCivilUnrest} \cite{sharkey_2020_perspective}. 

Now, we introduce our methodology for differentiating clusters within districts as being systemic vs non-systemic.

\subsection{Designing a Systemic Index: Procedure for defining a District with Systemic Crime vs Non-Systemic Crime  }

We want to differentiate systemic crime, where crime hotspots' geographic locations have not changed over time, from non-systemic crime, where they have changed significantly over time. In this sense, how 'systemic' the crime is corresponds to how much variation there is in the locations of clusters over the years. As it is quite complicated to determine variations in clusters, it would be prudent to instead represent these clusters by their \textit{cluster representative}, the average of their core points in DBSCAN. Formally, for a cluster $\mathcal{C}$, with $\mathcal{C}_{core}$ being the set of \textit{core points} in $\mathcal{C}$, the representative is
\begin{equation}
    \Tilde{x} = \underset{x \in \mathcal{C}_{core}}{\text{avg}} x .
\end{equation}Then, for each district, we have temporal data of these collections of \textit{cluster representatives}, and want to determine whether these sets change geospatially over time. The first step is to define a distance between two sets of points. 

Suppose we have two sets $S_1,S_2$ of points in the $2$D plane, and we want to see how 'far' apart they are. The traditional idea would be to use a permutation method. Here, we randomly permute the order of $S_2$, pair off points in $S_1$ and $S_2$, and get the sum of the distances between paired points. We average these sums over a large number of random permutations, and this will give a sort of distance metric on sets of points.

One major disadvantage of this method is its bias in classifying areas with multiple clusters as systemic crime, because the sum of distances will be larger. Also, it is not clear which points to pair off if the sizes do not match up. Furthermore, this metric is fairly inaccurate when two sets of points are close together: in fact, for two equal sets of \textit{sparse} points, the computed random permutation metric would be quite large, but the actual distance should be zero.

Thus, we propose a novel metric on two sets of points $S_1,S_2$. This is defined as follows, where $d$ represents the usual Euclidean metric between two points on a plane, with latitude and longitude being the coordinates:
\begin{equation}
\label{eqn:dmetric}
d_{set}(S_1,S_2) = \underset{p_1 \in S_1}{\text{avg}} \underset{p_2 \in S_2}{\text{min}} d(p_1,p_2) + \underset{p_2 \in S_2}{\text{avg}} \underset{p_1 \in S_1}{\text{min}} d(p_2,p_1).
\end{equation}

Essentially, for each point in $S_1$, we find the closest distance to a point in $S_2$, and then average all these minimal distances; we do the same process from each point in $S_2$ to $S_1$; and then finally sum these two averages. This avoids the bias of the random permutation metric, because it takes averages rather than sums. Furthermore, this handles the case where $|S_1| \neq |S_2|$ well, as it computes the average for each of the two sets and adds them. Finally, if two sets of points are close together, because we are taking an average of \textit{minimum distances}, then the metric will return a small value, which is fairly accurate.

With the above metric, we define the \textbf{non-systemic index} $\mathcal{NSI}(D)$ for each district $D$. Suppose we have an array of $n$ sets of points for the district $D$, say $[S_1,\ldots,S_n]$, where $S_i$ is the set of cluster representatives for year $i$. Our goal is to quantify the overall physical proximity of these sets. Indeed, by using the distance metric defined in Equation \ref{eqn:dmetric}, we can form an $n \times n$ \textit{distance matrix} $\mathcal{M}(D)$ for these sets, with entries
\begin{equation}
    (\mathcal{M}(D))_{i,j} = d_{set}(S_i,S_j), \; \; \; \forall \; 1 \leq i,j \leq n.
\end{equation}
Then, the $\mathcal{NSI}$ is computed as the Frobenius norm ($2$-norm) of this matrix:
\begin{equation}
    \mathcal{NSI}(D) = \|\mathcal{M}\|_2.
\end{equation}
Note that a higher $\mathcal{NSI}$ indicates that the distance matrix is larger in magnitude, meaning the cluster representatives physically move more over time, and thus the crime hotspots change significantly in that district. That is, a high $\mathcal{NSI}$ indicates that the district is \textbf{non-systemic}.

\begin{figure}[ht]
\begin{center}
\includegraphics[scale=0.55]{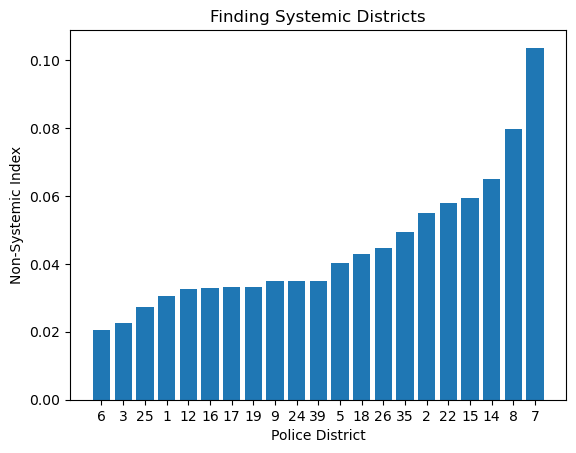}
\caption{Police districts are ranked in ascending order by non-systemic indices over the years 2012-2022.}
\label{fig:sys-metric-graph}
\end{center}
\end{figure} 

From the bar graph in Figure \ref{fig:sys-metric-graph}, we see that it will be interesting to study the underlying factors for the more systemic and non-systemic districts. Based on the $\mathcal{NSI}$, we will use \textbf{District 25 and District 19 in our analysis of systemic districts}, as they have extremely low scores. Also, we will study \textbf{District 7 and District 8 in our non-systemic analysis}, as they have the highest $\mathcal{NSI}$ values.

\subsection{Systemic Analysis}

We have defined systemic crime as crime in which clusters do not significantly change in geographic position over time. The goal is to now determine what underlying factor has caused the crime to remain in the same location. We propose and hypothesize that one of the major causes for this crime is attributed to systemic racism and the historical redlining that occurred in the city. 

\subsubsection{What is Redlining?}
Redlining, the practice of denying financial services such as loans or insurance to residents of certain neighborhoods based on their race or ethnicity, has had a significant impact on crime rates in affected communities. Redlining occurred primarily in the United States from the 1930s through the 1960s, although its legacy continues to affect many communities today \cite{hernandez_2022}. One of the ways in which redlining has impacted crime is by exacerbating poverty and economic inequality in redlined neighborhoods. By denying residents access to credit and other financial resources, redlining has made it more difficult for people in these neighborhoods to start businesses, buy homes, and otherwise build wealth \cite{fishback_2022}. This has led to higher levels of poverty and unemployment, which are both risk factors for criminal behavior \cite{aaronson_2021}.

\begin{figure}[ht]
\begin{center}
\includegraphics[width=0.8\linewidth]{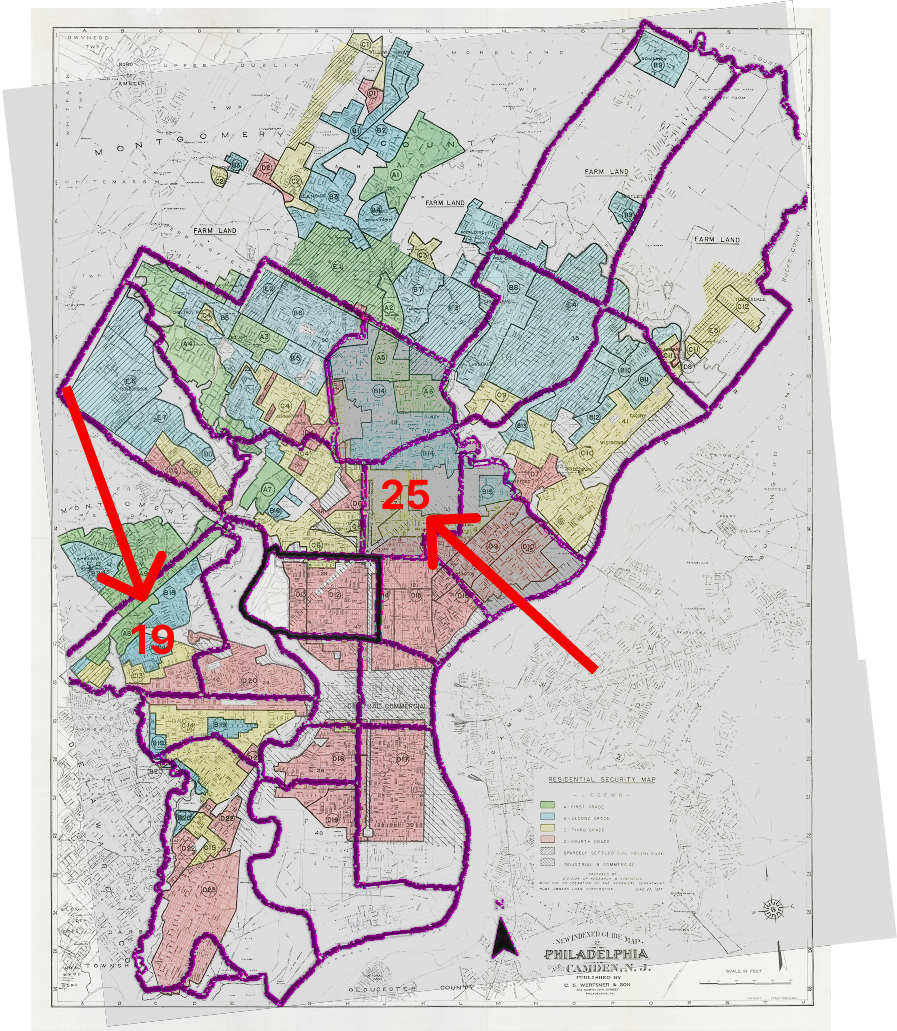}
\caption{Redlining Data from 1930s with Current Day Police District Map Overlayed}
\label{fig:redline-map}
\end{center}
\end{figure}

In Figure \ref{fig:redline-map}, we overlay the red-lined neighborhood geography on top of our clusters to visually examine whether there is an overlap in the red-lined neighborhoods and the clusters from DBScan. To support our hypothesis, we are correlating the presence of redlining and institutionalized racism. We map each latitude-longitude pair to redlining Home Owner's Loan Corporation (HOLC) grade (A is the best whereas D is the worst)- with Mapping Inequality data from The University of Richmond \cite{richmond}.

We elect to use police districts 19 ($\mathcal{NSI}=0.0332$) and 25 ($\mathcal{NSI}=0.0273$), as these are classified as systemic, falling below the median $\mathcal{NSI}$ for districts of $\approx 0.06$. For our analyses, we re-cluster each district with data across all years, as $\mathcal{NSI}$ established temporal homogeneity. We then compute the breakdown of HOLC grade for each cluster and observe notable differences and inequities. Both police districts are characterized by one far larger cluster and several relatively small clusters. We chose to combine the HOLC grade counts for these smaller clusters. To assure mitigation of introduced bias, we visually observe the HOLC grade breakdown from \ref{fig:redline-map}. By cross-referencing these with our earlier made DBSCAN clusters, we see, geographically, that the grades in smaller clusters are primarily homogeneous. We then use Chi Square Tests for Independence to determine if there is a statistically significant association between the two variables \cite{rice}.

\begin{table}[h]
\begin{tabular}{|c|c|c|}
\hline
HOLC Grade & Cluster 1 & Other Clusters \\
\hline
A                                                                            & 0.4\%                 & 7\%                       \\
\hline
B                                                                            & 20\%                  & 39\%                      \\
\hline
C                                                                            & 16\%                  & 52\%                      \\
\hline
D                                                                            & 63.6\%                & 2\%   \\                 \hline  
\end{tabular}
\caption{Percentage breakdown of HOLC grade by Cluster in District 19. ($\chi^2=90.2$, $p=2.02\times 10^{-19}$)}
\label{tab:my-table1}
\end{table}

\begin{table}[h]
\begin{tabular}{|c|c|c|}
\hline
HOLC Grade & Cluster 1 & Other Clusters \\
\hline
A                                                                            & 0\%                 & 0\%                       \\
\hline
B                                                                            & 17\%                  & 0\%                      \\
\hline
C                                                                            & 54\%                  & 100\%                      \\
\hline
D                                                                            & 29\%                & 0\%   \\                 \hline  
\end{tabular}
\caption{Percentage breakdown of HOLC grade by Cluster in District 25. ($\chi^2=59.0$, $p=1.52 \times 10^{-13}$)}
\label{tab:my-table2}
\end{table}

Both Districts 19 and 25 displayed significant associations with HOLC grade, implying differences between the two cluster groups. It is important to note that District 25 contained no "A" HOLC grade-regions and combined clusters were only in "C" grade regions, as detailed in Table \ref{tab:my-table2}. However, this still leads us to an important discussion surrounding the lasting effects of redlining.
\subsection{Non-Systemic Analysis}
We base our non-systemic analysis on police districts that displayed a high non-systemic index. This implies low variance in the geographic distribution of crime clusters from year to year. We attempt to determine if there is a present association between the clustered crime regions and the distribution of historical redlining grades.

As shown in Figure \ref{fig:sys-metric-graph}, the two districts that exhibited the most non-systemic behavior from 2012-2022 were district 7 and district 8. Let us now attempt to understand potential reasons why this observation may be arising. From district map data and online reading, we know that both districts 7 and 8 have large areas of farmland and are more agricultural/rural communities. Let us first turn to district 7, which, on average, has the lowest frequency of crime among all the districts (Figure \ref{fig:temp_dist}). As defined earlier, being non-systemic means having clusters of crime that change geographic locations over time. In general, given the smaller sample size (total number of crimes committed) in district 7, small changes in the number of crimes committed can have a significant impact on the overall crime rate and consequently the clustering.

Now, we can turn to district 8. Figure \ref{fig:temp_dist} shows that district 8 has the highest incidences of crime during the years 2014-2015. It is well known and well-documented that economic conditions can be a causal factor for crime \cite{fergusson_2004_how}. In particular, rural areas may be more vulnerable to changes in economic conditions, such as a downturn in the agricultural sector. When people lose their jobs or experience financial hardship, they may be more likely to engage in criminal behavior. News articles from 2014-2015 announced that the ``Average household farm income in 2015 is projected to be \$19,121, down from last year’s high of \$28,687, according to the USDA'' \cite{2015_farming}. Thus, the economic pressure in the agricultural communities of districts 7 and 8 may be leading to the non-system patterns we observe. The authors acknowledge that sometimes it can be easy to "explain-away" certain observations so further investigation into districts 7 and 8 would be necessary before making strong claims.

\subsection{Interpretation of Results: Targeted Intervention}

In our analysis, we have identified that police districts can be classified as either being hosts to systemic or non-systemic crime. This determination is extremely useful for legislators as data shows that different crime mitigation techniques are more effective within systemic vs non-systemic districts. Overall we can say that we have taken a major stride towards achieving our goal of creating actionable insights for effective/targeted crime mitigation strategies. 

While our model is capable of analyzing any particular district within the city of Philadelphia, we chose to analyze particular districts due to their more extreme $\mathcal{NSI}$ scores. As a recap, the $\mathcal{NSI}$  score allows us to determine the "systemic-ness" of a particular district. For the following districts, our model along with current event information could result in suggestions similar to the following.

Districts 7 and 8 (Northeast Philadelphia): The high $\mathcal{NSI}$ score indicates these districts as non-systemic. Poverty within this district has also recently been on the rise with neighborhoods such as the Northeast neighborhood of Holmesburg, which went from a 2 percent poverty rate to 19 percent \cite{miller_2022}.

Given these two factors, we suggest more reactionary measures. Specifically, a potential policy suggestion would be to increase police presence within this area as well as implementing economic stimulus legislation.

Our analyzed systemic Districts (19 and 25), both displayed significant statistical associations with HOLC grade. For District 19, we see that 63\% of crimes took place in regions historically rated as $D$. This is clearly vastly different from the 2\% in the other combined clusters. This implies that historical and institutionalized bias still plays an important role in determining the geographic distributions of crime. In District 25, the statistical results are not as reliable, as we see all smaller clusters reside in $C$ grade regions. However, the lack of $A$, and small amount of $B$ graded regions overall in District 25 leads us to a similar conclusion as District 19. In these cases, a push for such interventions as targeted economic development can attract investing and ultimately reduce poverty rates in this area. Another initiative that can combat the historical impacts of redlining, specifically in the context of crime, is community policing. Using such methods as DBSCAN, we have reliably identified crime hotspots within each District. With our research, police forces should formulate plans to increase attention towards these hotspots in the future. 
\subsection{Limitations}

While the Hopkins statistic can be a useful tool in determining the clustering tendency of a data set, there are some assumptions and limitations in its use. These include:

\begin{enumerate}

\item \textit{Sensitivity to data distribution:} the Hopkins statistic assumes that the data is uniformly distributed, and it may not work well if the data has a skewed distribution or contains outliers. It is unclear if the clustered data is normally distributed. 

\item \textit{The dataset has no noise:} The Hopkins statistic assumes that the dataset does not contain any noise or outliers that could bias the clustering tendency, but in the crime dataset there may be a large number of outliers in certain districts.

\item \textit{The dataset has a moderate to high number of dimensions:} The Hopkins statistic assumes that the dataset has a moderate to high number of dimensions, as otherwise, it might not be able to distinguish between a clustered and non-clustered dataset. Here, this is potentially problematic, as we only use two dimensions (latitude, longitude). \textit{Nevertheless,} a counterpoint is that because the dataset had many clusters and a complex structure, low dimensionality was not an issue.

\end{enumerate}
 Similarly, the $\chi^2$-model has a few assumptions and limitations to consider:

\begin{enumerate}
    \item \textit{Independence:} The observations should be ideally independent of each other here in the $\chi^2$-test. However, here, the HOLC grade observations may be dependent on each other, based on historical and demographic factors that have affected redlining.
    \item \textit{Sample size:} The sample size should be sufficiently large to ensure that the expected frequencies are not too low. This was not an issue in District 19, but was potentially problematic in District 25, where some clusters had zeroes in their percentage breakdown. 
    \item \textit{Association, not causation:} The $\chi^2$-model can detect an association between two variables, but it cannot determine whether this is causal.
\end{enumerate}
Violation of these assumptions can lead to incorrect results and conclusions. However, violating one or more of these assumptions does not necessarily mean that the Hopkins statistic or $\chi^2$-model cannot be used. Rather, it is generally important to be aware of these assumptions and their potential impact on the results of the analysis.

\section{Conclusion and Future Work}
In this paper, we began by exploring crime distributions within the city of Philadelphia and ultimately decided upon temporal and geospatial analysis within set police district boundaries. We then proposed a method for classifying the type of crime within these districts and began to explore the potential underlying causes including factors such as redlining. This research provides evidence that crime is not necessarily a uniform phenomenon. In this study, we found that crime can be differentiated by its temporal and geographical properties. Our results could allow policy makers to make more targeted interventions for the different types of crime. In order to further this work, we would like to explore different underlying factors and their relationships to the types of crime within the city. 


\bibliography{biblio}

\end{document}